# Quantifier Alternation in Two-Variable First-Order Logic with Successor Is Decidable

Manfred Kufleitner[*]   Alexander Lauser[*]

University of Stuttgart, FMI, Germany
{kufleitner,lauser}@fmi.uni-stuttgart.de

**Abstract.** We consider the quantifier alternation hierarchy within two-variable first-order logic $\mathrm{FO}^2[<,\mathrm{suc}]$ over finite words with linear order and binary successor predicate. We give a single identity of omega-terms for each level of this hierarchy. This shows that it is decidable for a given regular language and a nonnegative integer $m$, whether the language is definable by a formula in $\mathrm{FO}^2[<,\mathrm{suc}]$ which has at most $m$ quantifier alternations. We also consider the alternation hierarchy of unary temporal logic $\mathrm{TL}[\mathsf{X},\mathsf{F},\mathsf{Y},\mathsf{P}]$ defined by the maximal number of nested negations. This hierarchy coincides with the $\mathrm{FO}^2[<,\mathrm{suc}]$ alternation hierarchy.

**Keywords:** automata theory; semigroups; regular languages; first-order logic; finite model theory

## 1 Introduction

Around 1960, Büchi, Elgot and Trakhtenbrot independently showed that monadic second-order logic (MSO) over finite words defines exactly the class of regular languages [2, 6, 32]. Since then, numerous fragments of MSO have been considered. A theoretical motivation for fragments is the study of the rich structure within the regular languages. For this purpose, fragments form the basis of a descriptive complexity theory: The simpler the formula for defining a language, the simpler this language is. In contrast to classical complexity theory, in some cases one can actually check whether a given language lies in a certain descriptional complexity class. From a practical point of view, simpler fragments often lead to more efficient algorithms for decision problems such as satisfiability.

The most prominent fragment of MSO is first-order logic FO. The atomic predicates of FO are the unary predicate $\lambda(x) = a$ stating that position $x$ is labeled by the letter $a$, and the binary predicates $x = y$ and $x < y$ with the natural interpretation. The successor predicate $\mathrm{suc}(x,y)$ is easily definable in FO by saying that $x < y$ and that there is no position between $x$ and $y$. McNaughton and Papert showed that a language is FO-definable if and only if it is star-free [17]. Combined with Schützenberger's characterization of star-free languages in terms of finite aperiodic monoids [20], it follows that a language is FO-definable if and only if its syntactic monoid is aperiodic. The latter property is decidable and thus, one can effectively check whether a regular language (given *e.g.* by a nondeterministic automaton or an MSO formula) is definable in FO. The two most famous hierarchies within FO are the Straubing-Thérien hierarchy and Brzozowski's dot-depth hierarchy. The Straubing-Thérien hierarchy coincides with the quantifier alternation inside FO without the successor predicate [24, 28], and Brzozowski's dot-depth hierarchy is captured by quantifier alternation including the successor predicate [3]; see also [19, 30]. Here, quantifier alternation is defined in terms of blocks of quantifiers for formulae in prenex normal form. Note that by introducing new variables, every formula can be converted into prenex normal form. Deciding membership for level $m$ of one of the hierarchies for a given regular language is one of the most challenging open problems in automata theory. To date, for both hierarchies only the very first levels (*i.e.*, $m = 1$) are known to be decidable [9, 23].

[*]The authors acknowledge the support by the German Research Foundation (DFG) under grant DI 435/5-1.

By Kamp's Theorem [8], first-order logic $FO^3$ with only three different names for the variables and full first-order logic FO have the same expressive power. Two variables are not sufficient for defining all first-order definable languages; for example $(ac^*bc^*)^*$ is not definable in first-order logic with two variables $FO^2$, even in the presence of the successor predicate. The fragment $FO^2[<]$ without successor predicate has a huge number of different characterizations; see *e.g.* [5, 27]. One of them is the variety **DA** of finite monoids [21]; *cf.* [29]. For quantifier alternation inside $FO^2$ one cannot readily rely on prenex normal forms. However, in $FO^2$ negations can be moved towards the atomic formulae, and hence every formula is equivalent to a negation-free counterpart. The fragment $FO^2_m$ consists of all $FO^2$-formulae whose negation-free counterpart has at most $m$ blocks of quantifiers on each path of the parse tree. Kufleitner and Weil have shown that for every $m \geq 1$ it is decidable whether a given regular language is definable in $FO^2_m[<]$ without successor predicate [15]. They have given an effective algebraic characterization in terms of levels of the Trotter-Weil hierarchy of finite monoids [33]; see also [14]. In addition, restrictions of many other characterizations of the $FO^2[<]$-definable languages admit algebraic counterparts within this hierarchy [12, 16]. The proof of Kufleitner and Weil's characterization of $FO^2_m[<]$ relies on a combinatorial tool known under the terms *ranker* [34] and *turtle program* [22]. A connection between $FO^2_m[<]$ and rankers was established by Weis and Immerman [34] and further exploited by Kufleitner and Weil [16]. Straubing has given another algebraic characterization of $FO^2_m[<]$ in terms of weakly iterated block products of $\mathcal{J}$-trivial monoids [26]. Recently, Krebs and Straubing [10] were able to use this characterization for giving identities of omega-terms for $FO^2_m[<]$, thereby obtaining another effective characterization of $FO^2_m[<]$.

In this paper, we consider the quantifier alternation hierarchy inside $FO^2[<, \mathrm{suc}]$ with successor predicate. The logic $FO^2[<, \mathrm{suc}]$ is strictly more expressive than $FO^2[<]$ without successor. For example, the language $(ab)^*$ is definable in $FO^2[<, \mathrm{suc}]$ but not in $FO^2[<]$. Thérien and Wilke [29] gave an algebraic characterization of $FO^2[<, \mathrm{suc}]$ which, by a previous result of Almeida, is known to coincide with the decidable variety **LDA** of finite semigroups [1]; see also [4]. For every $m \geq 2$ we give a single identity of omega-terms such that a language is definable in $FO^2_m[<, \mathrm{suc}]$ if and only if its syntactic semigroup satisfies this identity. In particular, it is decidable whether a given regular language is $FO^2_m[<, \mathrm{suc}]$-definable.

Our proof is by induction on $m$ and the base case is Knast's Theorem on dot-depth one [9]. For $m = 1$, there is a small difference between the availability and the absence of min- and max-predicates; this is identical to the situation for dot-depth one [11]. The main ingredients of our proof are *(i)* string rewriting techniques, *(ii)* combinatorial properties of **LDA**, and *(iii)* relativization techniques for $\text{FO}^2$. As a byproduct, we show that quantifier alternation in $\text{FO}^2[<, \text{suc}]$ coincides with alternation in unary temporal logic $\text{TL}[\mathsf{X}, \mathsf{F}, \mathsf{Y}, \mathsf{P}]$ where the latter is based on the nesting depth of negations. This last property can also be seen using translations between $\text{FO}^2$ and unary temporal logic given by Etessami, Vardi, and Wilke [7].

## 2 Preliminaries

Throughout, $A$ denotes a finite alphabet. The set of all finite words is $A^*$ and the set of all finite, nonempty words is $A^+$. For $u = a_1 \cdots a_n$ with $a_i \in A$, the *length* of $u$ is $|u| = n$ and the *$k$-factor alphabet* is $\text{alph}_k(u) = \{a_i \cdots a_{i+k-1} \in A^k \mid 1 \leq i \leq n-k+1\}$. The set of *positions* of $u$ is $\text{pos}(u) = \{1, \ldots, n\}$. If $I$ is an interval, then $u[I]$ denotes the factor of $u$ covered by the interval of positions $\text{pos}(u) \cap I$. If $I = [i; j]$, then $u[i; j]$ is an abbreviation for $u[I]$. In particular, if $1 \leq i \leq j \leq n$, then $u[i; j] = a_i \cdots a_j$. If $i = j \in \text{pos}(u)$, then we simply write $u[i]$ to denote the $i^{\text{th}}$ letter of $u$.

**First-Order Logic and Unary Temporal Logic.** We consider first-order logic over finite words with order and successor predicates. Atomic first-order formulae are $\top$ for *true*, $\bot$ for *false*, label predicates $\lambda(x) = a$ for $a \in A$, comparisons $x = y$, $x < y$ and successor $\text{suc}(x, y)$ as well as minimum $\min(x)$ and maximum $\max(x)$. Words form models for formulae as labeled, linearly ordered sets of positions and $x$ and $y$ are variables ranging over positions. Formulae can be composed by the usual Boolean connectives, *i.e.*, if $\varphi$ and $\psi$ are first-order formulae, then so are their disjunction $\varphi \lor \psi$, their conjunction $\varphi \land \psi$, and the negation $\neg \varphi$. Moreover, formulae can be composed by existential quantification $\exists x \, \varphi$ and universal quantification $\forall x \, \varphi$. The semantics is as usual; see *e.g.* [13, 31]. We use the notation $\varphi(x_1, \ldots, x_n)$ to indicate that at most the variables $x_1, \ldots, x_n$ occur freely in $\varphi$. We write $u \models \varphi(i_1, \ldots, i_n)$ for $u \in A^*$ and positions $i_j \in \text{pos}(u)$ if $\varphi$ is true over $u$ with $x_j$ being interpreted by $i_j$. A formula without free variables is a *sentence* and in this case we simply write $u \models \varphi$. For any class $\mathcal{F}$ of first-order formulae, $\mathcal{F}[\mathcal{C}]$ is the restriction to formulae which, apart from $\top$, $\bot$, label predicates, and equality, only use predicates in $\mathcal{C} \subseteq \{<, \text{suc}, \min, \max\}$.

The fragment $\text{FO}^2 = \text{FO}^2[<, \text{suc}, \min, \max]$ of first-order logic contains all formulae which use at most two different names for variables, say $x$ and $y$. For $\text{FO}^2$-formulae $\varphi(x)$ with free variable $x$ we stipulate the convention that $\varphi(y)$ is the formula obtained by interchanging $x$ and $y$. Using De Morgan's laws and the usual dualities between existential and universal quantifiers, one can see that every formula in $\text{FO}^2$ is equivalent to a formula with negations only applied to atomic formulae. We call such formulae *negation-free* (since negative predicates could be added to an extended signature). The fragment $\text{FO}^2_m$ consists of all formulae in $\text{FO}^2$ with quantifier alternation depth at most $m$, *i.e.*, formulae such that the negation-free counterpart has at most $m$ blocks of quantifiers on every path of the parse tree. (Thus if we drop the two-variable restriction, every $\text{FO}^2_m$-formula admits a prenex normal form with $m$ blocks of quantifiers). In other words, negation-free formulae in $\text{FO}^2_m$ have at most $m - 1$ alternations of nested existential and universal quantifiers. Note that $\text{FO}^2_m$ is closed under negation. The fragment $\text{FO}^2_{m,n}$ contains all formulae in $\text{FO}^2_m$ with quantifier depth at most $n$.

Unary temporal logic TL[X, F, Y, P] consists of all formulae built from $\top$ for *true*, $\bot$ for *false*, labels $a$ for $a \in A$, compositions using Boolean connectives as in first-order logic, and temporal modalities $X\varphi$, $F\varphi$, $Y\varphi$, and $P\varphi$ for $\varphi \in$ TL[X, F, Y, P]. Formulae of unary temporal logic are interpreted over a word together with a position. The semantics is declared by the following FO$^2$-formulae in one free variable: Let $a(x) \equiv (\lambda(x) = a)$ and

$$(X\varphi)(x) \equiv \exists y \, (\text{suc}(x, y) \land \varphi(y)), \quad (F\varphi)(x) \equiv \exists y \, (x \leq y \land \varphi(y)),$$
$$(Y\varphi)(x) \equiv \exists y \, (\text{suc}(y, x) \land \varphi(y)), \quad (P\varphi)(x) \equiv \exists y \, (y \leq x \land \varphi(y)).$$

Here and in the sequel, $\equiv$ means syntactic equality. We often use this symbol instead of equality in order to avoid confusion with the symbol = occurring as atomic predicate. The formulae for the remaining constructs are as usual. The modalities X (neXt) and F (Future) are called *future modalities* whereas the modalities Y (Yesterday) and P (Past) are called *past modalities*. In order to define $u \models \varphi$ without a distinguished position in $u$, we start evaluation in front (position 0) for future modalities and after (position $|u| + 1$) the word $u$ for past modalities. More formally for a word $u \in A^*$, we define $u \not\models a$ and

$$u \models X\varphi \text{ if and only if } u \models \varphi(1), \quad u \models F\varphi \text{ if and only if } u \models F\varphi(1),$$
$$u \models Y\varphi \text{ if and only if } u \models \varphi(|u|), \quad u \models P\varphi \text{ if and only if } u \models P\varphi(|u|).$$

Boolean connectives and atomic formulae $\top$ and $\bot$ are defined as usual. For example, the formula $Xa \land Yb$ defines the language $aA^*b$. Let TL$_m$[X, F, Y, P] be the fragment of unary temporal logic consisting of the Boolean combinations of formulae with at most $m-1$ nested negations. Let TL$_{m,n}$[X, F, Y, P] consist of all formulae in TL$_m$[X, F, Y, P] with operator depth at most $n$, i.e., there are at most $n$ nested temporal modalities.

For a formula $\varphi$ in first-order logic or in unary temporal logic, let $L(\varphi) = \{u \in A^+ \mid u \models \varphi\}$ be the language defined by $\varphi$.

**Algebra.** Throughout this paper all semigroups are nonempty. Let $S$ be a finite semigroup. An element $x \in S$ is *idempotent* if $x^2 = x$. The set of all idempotents of $S$ is denoted $E(S)$. For every finite semigroup $S$ there exists an integer $\omega \geq 1$ such that each $\omega$-power is idempotent in $S$. *Green's relations* are an important concept in the structure theory of finite semigroups: For $x, y \in S$ let $x \leq_{\mathcal{R}} y$ if $x = y$ or $x \in yS$ and symmetrically let $x \leq_{\mathcal{L}} y$ if $x = y$ or $x \in Sy$. For $\mathcal{G} \in \{\mathcal{R}, \mathcal{L}\}$ let $x \, \mathcal{G} \, y$ if both $x \leq_{\mathcal{G}} y$ and $y \leq_{\mathcal{G}} x$; and let $x <_{\mathcal{G}} y$ if $x \leq_{\mathcal{G}} y$ but not $y \leq_{\mathcal{G}} x$. We also view $S$ as an alphabet and write $u \in S^+$ for a word with letters from $S$. For words $u, v \in S^+$ we say that a relation $u \, \mathcal{G} \, v$ "holds in $S$", if the relation is satisfied after evaluating $u$ and $v$ in $S$. We use this frequently for equality and Green's relations.

Frequently, classes of semigroups are defined by *identities* of omega-terms. An *omega-term* over a set of variables $\Sigma$ is defined inductively. Every $x \in \Sigma$ is an omega-term, and if $u$ and $v$ are omega-terms, then so are $uv$ and $u^\omega$. A finite semigroup $S$ *satisfies* the identity $u = v$ if for each homomorphism $h : \Sigma^+ \to S$ we have $h(u) = h(v)$. Here, $h$ is extended to omega-terms by letting $h(u^\omega)$ be the idempotent generated by $h(u)$.

Next, we define classes $\mathbf{W}_m$ of finite semigroups. To this end, we inductively define sequences of omega-terms $U_m, V_m$ with variables $e, f, x_i, y_i, s, t, p_i, q_i$. For $m \geq 2$ we let

$$U_1 = (e^\omega s f^\omega x_1 e^\omega)^\omega s (f^\omega y_1 e^\omega t f^\omega)^\omega, \quad U_m = (p_m U_{m-1} q_m x_m)^\omega p_m U_{m-1} q_m (y_m p_m U_{m-1} q_m)^\omega,$$
$$V_1 = (e^\omega s f^\omega x_1 e^\omega)^\omega t (f^\omega y_1 e^\omega t f^\omega)^\omega, \quad V_m = (p_m U_{m-1} q_m x_m)^\omega p_m V_{m-1} q_m (y_m p_m U_{m-1} q_m)^\omega.$$

By definition, a semigroup is in $\mathbf{W}_m$ if it satisfies the identity $U_m = V_m$. The class $\mathbf{W}_1$ is Knast's algebraic characterization of dot-depth one [9]. The only difference between $U_1$ and $V_1$ is the central variable in $U_1$ being $s$ and in $V_1$ being $t$. Intuitively, this difference is more and more "shielded" in $U_m$ and $V_m$ with increasing $m$.

For every $e \in E(S)$ the set $eSe$ forms the so-called *local monoid* at $e$. A semigroup $S$ belongs to **LDA** if every local monoid $eSe$ satisfies $(xy)^\omega x(xy)^\omega = (xy)^\omega$. This is equivalent to saying that we have $(exeye)^\omega exe(exeye)^\omega = (exeye)^\omega$ in $S$ for all $x, y \in S$ and all $e \in E(S)$. Note that if $S$ is in **LDA** and if $e \in E(S)$ and $x, y \in eSe$ then,

$$(xy)^\omega = (xy)^{\omega-1}x(yx)^\omega y = (xy)^{\omega-1}x(yx)^\omega y(yx)^\omega y = (xy)^{2\omega}y(xy)^\omega = (xy)^\omega y(xy)^\omega.$$

Thus despite its asymmetric definition, **LDA** is left-right-symmetric.

A homomorphism $h : A^+ \to S$ to a finite semigroup $S$ *recognizes* a language $L \subseteq A^+$ if $h^{-1}(h(L)) = L$. A semigroup $S$ *recognizes* $L \subseteq A^+$ if there exists a homomorphism $h : A^+ \to S$ which recognizes $L$. For $u, v \in A^+$ let $u \equiv_L v$ if $puq \in L$ is equivalent to $pvq \in L$ for all $p, q \in A^*$. The relation $\equiv_L$ over $A^+$ is a congruence and the semigroup $A^+/\equiv_L$, also denoted by $\mathrm{Synt}(L)$ and called the *syntactic semigroup* of $L$, is the unique minimal semigroup recognizing $L$. Moreover, it is effectively computable (*e.g.* from an automaton for $L$); *cf.* [18].

## 3 Alternation within Two-Variable First-Order Logic with Successor

The following result is the main contribution of this paper. The remainder of this section is dedicated to its proof.

**Theorem 1** *Let $m \geq 2$ and let $L \subseteq A^+$. The following assertions are equivalent:*
*1) $L$ is definable in $\mathrm{FO}_m^2[<, \mathrm{suc}]$.*
*2) $L$ is definable in $\mathrm{TL}_m[\mathsf{X}, \mathsf{F}, \mathsf{Y}, \mathsf{P}]$.*
*3) $\mathrm{Synt}(L) \in \mathbf{W}_m$.*

Before turning to the proof of Theorem 1 we record the following decidability corollary. For $m = 1$ it relies on a characterization of two-sided ideals inside dot-depth one [11].

**Corollary 1** *For every positive integer $m$ one can decide whether a given regular language $L \subseteq A^+$ is definable in $\mathrm{FO}_m^2[<, \mathrm{suc}]$.*

*Proof:* By Theorem 1, it remains to prove the case $m = 1$. A language is definable in $\mathrm{FO}_1^2[<, \mathrm{suc}]$ if and only if it is Boolean combination of languages of the form $A^* w_1 A^* \cdots w_\ell A^*$. This is known to be equivalent to $\mathrm{Synt}(L) \in \mathbf{W}_1$ and the image of $L$ under the syntactic homomorphism being a union of so-called $\mathcal{J}$-classes, *cf.* [11]. Both properties can be verified effectively. □

In Subsection 3.1 we start with the implication from (3) to (1) in Theorem 1. This is essentially Proposition 1 whose proof requires some preparatory work: We first show that every $\mathbf{W}_m$ is contained in **LDA** (Lemma 1) which enables us to use a combinatorial property of **LDA** (given in Lemma 4). Then a relativization technique for $\mathrm{FO}_m^2$ (Lemma 5) is used for defining a congruence $\approx_{m,n}$ (Definition 1) as a tool for $\mathrm{FO}_m^2$. The connection between this congruence and $\text{FO}^2_m$ is established by Lemma 7. Using a string rewriting system, a special factorization (given in Lemma 9) finally leads to an inductive scheme to prove Proposition 1.

In Subsection 3.2 the reverse implication from (1) to (3) is given by Lemma 11. It is shown that if a language is definable in $\text{FO}^2_m[<, \text{suc}]$, then it is recognized by a semigroup in $\mathbf{W}_m$. Since the syntactic semigroup is a divisor of every recognizing semigroup, it follows that the syntactic semigroup of the language is in $\mathbf{W}_m$. Lemma 10 is an intermediate result which shows how to reduce the alternation depth by encoding some information into an extended alphabet.

Finally in Subsection 3.3, Proposition 2 incorporates unary temporal logic into the picture.

## 3.1 From Algebra to First-Order Logic

**Lemma 1** *For all $m \geq 1$ we have $\mathbf{W}_m \subseteq \mathbf{LDA}$.*

*Proof:* Let $S$ be a finite semigroup and let $\omega \geq 1$ be an integer such that $x^\omega$ is idempotent for all $x \in S$. Let $x, y \in S$ and let $e \in S$ be idempotent. Setting $e_1 = f_1 = s = e$, $x_1 = xey$, $y_1 = x$, $t = y$ we get $U_1 = (exeye)^\omega$ in $S$ and $V_1 = (exeye)^\omega eye(exeye)^\omega$ in $S$. Setting all other variables occurring in $U_m$ or in $V_m$ to be $e$, we see $U_m = (exeye)^\omega$ in $S$ and $V_m = (exeye)^\omega eye(exeye)^\omega$ in $S$. Thus if $S \in \mathbf{W}_m$ and $e \in E(S)$, then $eSe$ satisfies the identity $(xy)^\omega = (xy)^\omega y(xy)^\omega$, i.e., $S \in \mathbf{LDA}$. □

The next lemma is an intermediate result for Lemma 3 and Lemma 4 both of which yield important combinatorial properties of semigroups in $\mathbf{LDA}$.

**Lemma 2** *Let $S \in \mathbf{LDA}$, let $x, y, z \in S$, and let $e \in E(S)$.*
*1) If $xe \mathcal{R} ye$ in $S$, then $xe \mathcal{R} xez$ if and only if $ye \mathcal{R} yez$.*
*2) If $ex \mathcal{L} ey$ in $S$, then $ex \mathcal{L} zex$ if and only if $ey \mathcal{L} zey$.*

*Proof:* Since $\mathbf{LDA}$ is left-right symmetric, it suffices to show (1). Suppose $xe \mathcal{R} xez$. Since $ye \mathcal{R} xe \mathcal{R} xez$ there exist $s, t$ such that $xe = yes$ and $ye = xezt$. We get $ye = ye(esezte)$. Pumping the factor in the parentheses and using $\mathbf{LDA}$ yields $ye = ye(esezte)^\omega = ye(esezte)^\omega ezte(esezte)^\omega \in yezS$. □

**Lemma 3** *Let $S \in \mathbf{LDA}$, let $u, v \in S^+$, let $s, t \in S^*$, let $|v| \geq |S|$ and let $\text{alph}_{|S|+1}(vs) = \text{alph}_{|S|+1}(vt)$.*
*1) If $u \mathcal{R} uv$ in $S$, then $u \mathcal{R} uvs$ in $S$ if and only if $u \mathcal{R} uvt$ in $S$.*
*2) If $u \mathcal{L} vu$ in $S$, then $u \mathcal{L} svu$ in $S$ if and only if $u \mathcal{L} tvu$ in $S$.*

*Proof:* Since $\mathbf{LDA}$ is left-right symmetric, it suffices to show (1). Assume $u \mathcal{R} uv \mathcal{R} uvs$ in $S$. We want to show $u \mathcal{R} uvt$ in $S$. This is trivial if $t$ is the empty word. Otherwise we factorize $vt = pwz$ such that $|w| < |wz| = |S| + 1$ with $w = we$ in $S$ for some idempotent $e$ of $S$. Note that every sequence $x_1, \ldots, x_{|S|} \in S$ has a prefix which admits an idempotent stabilizer, i.e., there exists $i \in \{1, \ldots, |S|\}$ and $e \in E(S)$ such that $x_1 \cdots x_i = x_1 \cdots x_i e$ in $S$; see *e.g.* [11, Lemma 1] for a proof of this claim. Since $vs$ and $vt$ have the same factors of length $|S| + 1$, we find a factorization $vs = s_1 w z s_2$. Let $x = us_1 w$ and $y = upw$. By induction $u \mathcal{R} y$ and thus $xe = x \mathcal{R} y = ye$ in $S$. Moreover, $xe \mathcal{R} xez$ and by Lemma 2 we see $ye \mathcal{R} yez$ in $S$. This implies the claim. □

Choosing $s$ to be the empty word and $t = a$ immediately yields the following consequence.

**Lemma 4** *Let $S \in \mathbf{LDA}$, let $u, v \in S^+$, let $a \in S$ and let $|v| \geq |S|$.*
*1) If $u \mathcal{R} uv >_\mathcal{R} uva$ in $S$, then $\mathrm{alph}_{|S|+1}(v) \neq \mathrm{alph}_{|S|+1}(va)$.*
*2) If $u \mathcal{L} vu >_\mathcal{L} avu$ in $S$, then $\mathrm{alph}_{|S|+1}(v) \neq \mathrm{alph}_{|S|+1}(av)$.* $\square$

The next lemma gives the main combinatorial properties of $\mathrm{FO}^2_m[<, \mathrm{suc}]$ for our purpose, namely relativizations of formulae to certain factors of deterministic factorizations.

**Lemma 5** *Let $\varphi \in \mathrm{FO}^2[<, \mathrm{suc}]$ and let $v, w \in A^+$.*
*1) There exist formulae $\langle \varphi \rangle_{<\mathsf{X}w}$ and $\langle \varphi \rangle_{>\mathsf{X}w}$ such that for all $u = u_1 w u_2$ with a unique occurrence of the factor $w$ in the prefix $u_1 w$ we have*

$$u \models \langle \varphi \rangle_{<\mathsf{X}w}(i,j) \quad \text{iff} \quad u_1 \models \varphi(i,j) \quad \text{for all } 1 \leq i, j \leq |u_1|,$$
$$u \models \langle \varphi \rangle_{>\mathsf{X}w}(i,j) \quad \text{iff} \quad u_2 \models \varphi(i - |u_1 w|, j - |u_1 w|) \quad \text{for all } |u_1 w| < i, j \leq |u|.$$

*2) There exist formulae $\langle \varphi \rangle_{<\mathsf{Y}v}$ and $\langle \varphi \rangle_{>\mathsf{Y}v}$ such that for all $u = u_1 v u_2$ with a unique occurrence of the factor $v$ in the suffix $v u_2$ we have*

$$u \models \langle \varphi \rangle_{<\mathsf{Y}v}(i,j) \quad \text{iff} \quad u_1 \models \varphi(i,j) \quad \text{for all } 1 \leq i, j \leq |u_1|,$$
$$u \models \langle \varphi \rangle_{>\mathsf{Y}v}(i,j) \quad \text{iff} \quad u_2 \models \varphi(i - |u_1 v|, j - |u_1 v|) \quad \text{for all } |u_1 v| < i, j \leq |u|.$$

*3) There exists a formula $\langle \varphi \rangle_{[v;w]}$ such that for all $u = u_1 v u_2 w u_3$ with a unique occurrence of the factor $v$ in the suffix $v u_2 w u_3$ and a unique occurrence of the factor $w$ in the prefix $u_1 v u_2 w$ we have*

$$u \models \langle \varphi \rangle_{[v;w]}(i,j) \quad \text{iff} \quad u_2 \models \varphi(i - |u_1 v|, j - |u_1 v|) \quad \text{for all } |u_1 v| < i, j \leq |u_1 v u_2|.$$

*Moreover, if $\varphi \in \mathrm{FO}^2_{m,n}[<, \mathrm{suc}]$, then*
*1) $\langle \varphi \rangle_{<\mathsf{X}w} \in \mathrm{FO}^2_{m+1, n+|w|}[<, \mathrm{suc}]$ and $\langle \varphi \rangle_{>\mathsf{X}w} \in \mathrm{FO}^2_{m, n+|w|}[<, \mathrm{suc}]$,*
*2) $\langle \varphi \rangle_{<\mathsf{Y}v} \in \mathrm{FO}^2_{m, n+|v|}[<, \mathrm{suc}]$ and $\langle \varphi \rangle_{>\mathsf{Y}v} \in \mathrm{FO}^2_{m+1, n+|v|}[<, \mathrm{suc}]$, and*
*3) $\langle \varphi \rangle_{[v;w]} \in \mathrm{FO}^2_{m+1, n+N}[<, \mathrm{suc}]$ for $N = \max\{|v|, |w|\}$.*

*Proof:* The properties of the factorizations of $u$ mean that we actually consider the first occurrence of the factor $w$ (the last occurrence of the factor $v$, respectively). The letter $\mathsf{X}$ and $\mathsf{Y}$ derive from modalities of unary temporal logic where $\mathsf{X}_w$ is for neXt-$w$ which goes to the first occurrence of $w$, and $\mathsf{Y}_v$ is for Yesterday-$v$ which goes to the last occurrence of $v$.

"(1)": The idea for $\langle \varphi \rangle_{<\mathsf{X}w}$ and $\langle \varphi \rangle_{>\mathsf{X}w}$ is to restrict the interpretation of a formula to the part of the model which lies before (respectively after) the first occurrence of $w$. The construction is by structural induction. We first construct the formula $\langle \varphi \rangle_{<\mathsf{X}w}$. Let $\langle \varphi \rangle_{<\mathsf{X}w} \equiv \varphi$ if $\varphi$ is an atomic formula. Disjunction is given by $\langle \varphi \vee \psi \rangle_{<\mathsf{X}w} \equiv \langle \varphi \rangle_{<\mathsf{X}w} \vee \langle \psi \rangle_{<\mathsf{X}w}$, conjunction by $\langle \varphi \wedge \psi \rangle_{<\mathsf{X}w} \equiv \langle \varphi \rangle_{<\mathsf{X}w} \wedge \langle \psi \rangle_{<\mathsf{X}w}$, and negation by $\langle \neg \varphi \rangle_{<\mathsf{X}w} \equiv \neg \langle \varphi \rangle_{<\mathsf{X}w}$. For existential quantification we let

$$\langle \exists x\, \varphi \rangle_{<\mathsf{X}w} \equiv \exists x \left( (\neg \exists y \leq x \colon \lambda(y) = \vec{w}) \wedge \langle \varphi \rangle_{<\mathsf{X}w} \right).$$

Here $\lambda(y) = \vec{w}$ requires that the factor $w$ begin at position $y$. More formally, let $\lambda(y) = \vec{w}$ be defined as $\lambda(y) = w$ if $w \in A$ and as $\lambda(y) = a \wedge (\exists x \colon \mathrm{suc}(y, x) \wedge \lambda(x) = \vec{q})$ if $w = aq$ for $a \in A$ and $q \in A^+$. Note that the formula $\lambda(y) = \vec{w}$ is a purely existential $\mathrm{FO}^2$-formula with quantifier depth $|w| - 1$. Universal quantification is given dually by $\langle \forall x\, \varphi \rangle_{<\mathsf{X}w} \equiv \neg \langle \exists x\, \neg \varphi \rangle_{<\mathsf{X}w}$. As usual, swapping the variables $x$ and $y$ yields the corresponding constructions for $y$.

Next, we give the construction of $\langle \varphi \rangle_{>\mathsf{X}w}$. Let $\langle \varphi \rangle_{>\mathsf{X}w} \equiv \varphi$ if $\varphi$ is an atomic formula. Boolean combinations are similar to the above. For existential quantification let

$$\langle \exists x\, \varphi \rangle_{>\mathsf{X}w} \equiv \exists x \left( (\exists y < x \colon \lambda(y) = \overleftarrow{w}) \wedge \langle \varphi \rangle_{>\mathsf{X}w} \right).$$

Here $\lambda(y) = \overleftarrow{w}$ states that the factor $w$ ends at position $y$; it is defined left-right symmetrically to $\lambda(y) = \vec{w}$. Again, universal quantification and the constructions for $y$ are dual.

"(2)": This is left-right symmetric to the above.

Finally, we give the construction for $\langle \varphi \rangle_{[v;w]}$. Atomic formulae and Boolean combinations are defined as before. For existential quantification we set

$$\langle \exists x\, \varphi \rangle_{[v;w]} \equiv \exists x \left( \neg (\exists y \leq x \colon \lambda(y) = \vec{w}) \wedge \neg (\exists y \geq x \colon \lambda(y) = \overleftarrow{v}) \wedge \langle \varphi \rangle_{[v;w]} \right).$$

Quantification over $y$ and universal quantification are dual. $\square$

The relativization of the previous lemma leads to the congruence in the following definition. This congruence is our tool for the combinatorics of $\mathrm{FO}^2_m$ in the subsequent proofs.

**Definition 1** *Let $u, v \in A^*$. For $m, n \geq 0$ we let $u \approx_{m,0} v$ and $u \approx_{0,n} v$. For $n \geq 1$ let $u \approx_{1,n} v$ if $u$ and $v$ are contained in the same monomials $w_1 A^+ w_2 \cdots A^+ w_\ell$ with $w_i \in A^+$ and $|w_1 \cdots w_\ell| \leq n$. For $m \geq 2$ and $n \geq 1$ let $u \approx_{m,n} v$ if $\mathrm{alph}_k(u) = \mathrm{alph}_k(v)$ and $\mathrm{pref}_k(u) = \mathrm{pref}_k(v)$ and $\mathrm{suff}_k(u) = \mathrm{suff}_k(v)$ for all $k \leq n$, and all of the following hold:*

1) *if $u = u_1 w u_2$ and $v = v_1 w v_2$ with $1 \leq |w| \leq n$ such that the factor $w$ has a unique occurrence in the prefixes $u_1 w$ and $v_1 w$, then $u_1 \approx_{m-1, n-|w|} v_1$ and $u_2 \approx_{m, n-|w|} v_2$,*

2) *if $u = u_1 w u_2$ and $v = v_1 w v_2$ with $1 \leq |w| \leq n$ such that the factor $w$ has a unique occurrence in the suffixes $w u_2$ and $w v_2$, then $u_1 \approx_{m, n-|w|} v_1$ and $u_2 \approx_{m-1, n-|w|} v_2$.*

3) *if $u = u_1 w u_2 w' u_3$ and $v = v_1 w v_2 w' v_3$ with $|w w'| \leq n$ such that the factor $w$ has a unique occurrence in the suffixes $w u_2 w' u_3$ and $w v_2 w' v_3$ and such that the factor $w'$ has a unique occurrence in the prefixes $u_1 w u_2 w'$ and $v_1 w v_2 w'$, then $u_2 \approx_{m-1, n-|ww'|} v_2$.*

An elementary verification shows that $\approx_{m,n}$ is a congruence. Since this is not used in this paper, we do not record this fact as lemma.

**Lemma 6** *For all $u, v \in A^*$ and $m, n \geq 1$ we have that $u \approx_{m,n} v$ implies $u \approx_{m-1,n} v$ and $u \approx_{m,n-1} v$.*

*Proof:* The second claim is obvious. The first claim is also clear if $m = 1$ or if $m \geq 3$. It remains to consider the case $m = 2$. Let $u \approx_{2,n} v$. We shall show that $u \in P$ implies $v \in P$ for all $P = A^+ w_1 \cdots A^+ w_\ell$ with $w_i \in A^+$ such that $|w_1 \cdots w_\ell| \leq n$. This is true if $\ell = 1$ because $\mathrm{suff}_{|w_\ell|}(u) = \mathrm{suff}_{|w_\ell|}(v)$. Let $u = u_1 w_1 u_2$ and $v = v_1 w_1 v_2$ such that in both prefixes $u_1 w_1$ and $v_1 w_1$ there is a unique occurrence of the factor $w_1$, i.e., consider the first $w_1$ of $u$ and of $v$. Note that since $\mathrm{alph}_{|w_1|}(u) = \mathrm{alph}_{|w_1|}(v)$ such a factorization of $v$ exists. By definition, $u_2 \approx_{2, n-|w_1|} v_2$. Moreover, $u_2 \in A^+ w_2 \cdots A^+ w_\ell$ and by induction $v_2 \in A^+ w_2 \cdots A^+ w_\ell$. Moreover, $v_1 \in A^+$ because $\mathrm{pref}_{|w_1|}(u) = \mathrm{pref}_{|w_1|}(v)$ and thus $v \in P$. Now, if $u \in w_1 A^+ w_2 \cdots A^+ w_\ell$ with $u = w_1 u'$, then $v = w_1 v'$ and by definition $u' \approx_{2, n-|w_1|} v'$. The above claim yields $v' \in A^+ w_2 \cdots A^+ w_\ell$, i.e., $v \in w_1 A^+ w_2 \cdots A^+ w_\ell$. Symmetry yields $u \approx_{1,n} v$. $\square$

The next lemma connects $\mathrm{FO}^2_{m,n}$ with the combinatorial properties captured by $\approx_{m,n}$. For $u,v \in A^*$ let $u \equiv_{1,n} v$ if $u$ and $v$ model the same formulae in $\mathrm{FO}^2_{1,n}[<, \mathrm{suc}, \mathrm{min}, \mathrm{max}]$. For $m \geq 2$ and $u, v \in A^*$ let $u \equiv_{m,n} v$ if $u$ and $v$ model the same formulae in $\mathrm{FO}^2_{m,n}[<, \mathrm{suc}]$. For technical reasons, we have to include min and max predicates at level 1.

**Lemma 7** *For all $m, n \geq 0$ we have that $u \equiv_{m,n+1} v$ implies $u \approx_{m,n} v$.*

*Proof:* To prove the claim, we give a shortcut which extends the label predicate. Let $w_i \in A^+$ and $\psi(x)$ be a $\mathrm{FO}^2$-formula with free variable $x$. The formula $\lambda(x) = (\vec{w}_1, \ldots, \vec{w}_\ell, \psi)$ with free variable $x$ requires that, starting from the position $x$, there be occurrences of the factors $w_1, \ldots, w_\ell$ in this order such that on the last position $i$ of the occurrence of $w_\ell$ the formula $\psi(i)$ hold. More precisely, for every $u \in A^+$ and every $x \in \mathrm{pos}(u)$ we have $u \models (\lambda(x) = (\vec{w}_1, \ldots, \vec{w}_\ell, \psi))$ if and only if there exist $v_\ell \in A^*$ and $v_i \in A^+$, $1 \leq i < \ell$, such that $u[x; |u|] = w_1 v_1 \cdots w_\ell v_\ell$ and $u \models \psi(i)$ for $i = x + |w_1 v_1 \cdots w_\ell| - 1$. Formally, for $w = aw'$ with $w' \in A^*$ we define $\lambda(x) = (\vec{w}, \psi)$ by

$$\lambda(x) = a \wedge \exists y \colon (\mathrm{suc}(x,y) \wedge \lambda(y) = (\vec{w}', \psi)) \quad \text{if } |w'| > 0,$$
$$\lambda(x) = a \wedge \psi(x) \quad \text{else.}$$

We extend this notation inductively to sequences of factors by setting $\lambda(x) = (\vec{w}_1, \ldots, \vec{w}_\ell, \psi)$ to $\lambda(x) = (\vec{w}_1, \psi')$ where $\psi'(x) \equiv \exists y > x \colon \neg \mathrm{suc}(x,y) \wedge (\lambda(y) = (\vec{w}_2, \ldots, \vec{w}_\ell, \psi(y)))$. As usual, $\psi(y)$ is obtained from $\psi(x)$ by interchanging the variables $x$ and $y$. If $\ell = 1$ and $\psi = \top$, then we simply write $\lambda(x) = \vec{w}_1$.

Suppose $m = 1$ and $n \geq 1$. Let $P = w_1 A^+ w_2 \cdots A^+ w_\ell$ with $w_i \in A^+$ and $|w_1 \cdots w_\ell| \leq n$. Then the formula

$$\varphi \equiv \exists x \colon \mathrm{min}(x) \wedge (\lambda(x) = (\vec{w}_1, \ldots, \vec{w}_\ell, \mathrm{max}(x))).$$

is in $\mathrm{FO}^2_{1,n}[<, \mathrm{suc}, \mathrm{min}, \mathrm{max}]$. We have $u \in P$ if and only if $u \models \varphi$ if and only if $v \models \varphi$ if and only if $v \in P$. The second equivalence holds by assumption.

Suppose $m \geq 2$, $n \geq 1$ and $u \equiv_{m,n+1} v$. Let

$$\varphi_1 \equiv \exists x \colon (\lambda(x) = (\vec{w}, \top)),$$
$$\varphi_2 \equiv \exists x \colon (\lambda(x) = (\vec{w}, \top)) \wedge (\forall y \colon x \leq y),$$
$$\varphi_3 \equiv \exists x \colon (\lambda(x) = (\vec{w}, \forall y \colon y \leq x)),$$

for $w \in A^+$ with $|w| \leq n$. Then by assumption $u \models \varphi_i$ if and only if $v \models \varphi_i$. This respectively yields $\mathrm{alph}_k(u) = \mathrm{alph}_k(v)$, $\mathrm{pref}_k(u) = \mathrm{pref}_k(v)$, and $\mathrm{suff}_k(u) = \mathrm{suff}_k(v)$ for all $k \leq n$. Note that $\varphi_2$ and $\varphi_3$ may have quantifier depth $n + 1$. It remains to show Definition 1 (1–3).

Definition 1 (1): Let $u = u_1 w u_2$ and $v = v_1 w v_2$ with $1 \leq |w| \leq n$ such that the factor $w$ has a unique occurrence in the prefixes $u_1 w$ and $v_1 w$. First, we show $u_2 \approx_{m,n-|w|} v_2$. Using the assumption and Lemma 5, we see that for all $\varphi \in \mathrm{FO}^2_{m,n-|w|+1}[<, \mathrm{suc}]$ we have $u_2 \models \varphi$ if and only if $u \models \langle \varphi \rangle_{>\mathsf{X}w}$ if and only if $v \models \langle \varphi \rangle_{>\mathsf{X}w}$ if and only if $v_2 \models \varphi$. Note that $\langle \varphi \rangle_{>\mathsf{X}w} \in \mathrm{FO}^2_{m,n+1}[<, \mathrm{suc}]$. Thus $u_2 \equiv_{m,n-|w|+1} v_2$ and induction on $n$ yields $u_2 \approx_{m,n-|w|} v_2$.

Next, we show $u_1 \approx_{m-1,n-|w|} v_1$. Suppose $m = 2$ and let $P = w_1 A^+ w_2 \cdots A^+ w_\ell$ with $|w_1 \cdots w_\ell| \leq n - |w|$. Let

$$\psi(x) \equiv (\exists y \colon \mathrm{suc}(x,y) \wedge \lambda(y) = \vec{w}) \wedge \neg(\exists y < x \colon \lambda(y) = \vec{w}) \quad \text{and}$$
$$\varphi \equiv \exists x \colon (\forall y \colon x \leq y) \wedge \lambda(\vec{w}_1, \ldots, \vec{w}_\ell, \psi).$$

Note that $\varphi \in \mathrm{FO}^2_{2,n}[<,\mathrm{suc}]$. By assumption we get $u_1 \in P$ if and only if $u \models \varphi$ if and only if $v \models \varphi$ if and only if $v_1 \in P$. This shows $u_1 \approx_{1,n-|w|} v_1$. Suppose now $m \geq 3$. Using the formulae $\langle\varphi\rangle_{<\mathsf{X}_w} \in \mathrm{FO}^2_{m,n+1}[<,\mathrm{suc}]$ from Lemma 5, we see $u_1 \models \varphi$ if and only if $u \models \langle\varphi\rangle_{<\mathsf{X}_w}$ if and only if $v \models \langle\varphi\rangle_{<\mathsf{X}_w}$ if and only if $u_1 \models \varphi$ for all $\varphi \in \mathrm{FO}^2_{m-1,n-|w|+1}[<,\mathrm{suc}]$. Thus $u_1 \equiv_{m-1,n-|w|+1} v_1$ and inductively $u_1 \approx_{m-1,n-|w|} v_1$.

Definition 1 (2): This is left-right symmetric to the above.

Definition 1 (3): Let $u = u_1 w u_2 w' u_3$ and $v = v_1 w v_2 w' v_3$ with $|ww'| \leq n$ such that the factor $w$ has a unique occurrence in the suffixes $wu_2 w' u_3$ and $wv_2 w' v_3$ and such that the factor $w'$ has a unique occurrence in the prefixes $u_1 w u_2 w'$ and $v_1 w v_2 w'$. First suppose $m = 2$ and let $P = w_1 A^+ w_2 \cdots A^+ w_\ell$ with $|w_1 \cdots w_\ell| \leq n - |ww'|$. Let

$$\psi(x) \equiv (\exists y\colon \mathrm{suc}(x,y) \wedge \lambda(y) = \vec{w}) \wedge (\neg \exists y \leq x\colon \lambda(y) = \vec{w}) \quad \text{and}$$
$$\varphi \equiv \exists x\colon (\exists y\colon \mathrm{suc}(y,x) \wedge \lambda(y) = \vec{\bar{w}}) \wedge (\neg \exists y \geq x\colon \lambda(y) = \vec{\bar{w}}) \wedge \big(\lambda(x) = (w_1,\ldots,w_\ell,\psi)\big).$$

Note that $\varphi \in \mathrm{FO}^2_{2,n}[<,\mathrm{suc}]$. We have $u_2 \in P$ if and only if $u \in \varphi$ if and only if $v \in \varphi$ if and only if $v_2 \in P$ where the second equivalence holds by assumption. This shows $u_2 \approx_{1,n-|ww'|} v_2$. For $m \geq 3$ using the relativization of Lemma 5 (3) yields that $u \equiv_{m,n+1} v$ implies $u_2 \equiv_{m-1,n-|ww'|+1} v_2$. Induction thus yields $u_2 \approx_{m-1,n-|ww'|} v_2$. □

In other words the previous lemma shows that $\equiv_{m,n+1}$ is a refinement of $\approx_{m,n}$. In particular, $\approx_{m,n}$ has finite index. The next lemma essentially is an auxiliary statement used in the proof of Lemma 9. It says that $\approx_{1,n}$ equivalence of $u$ and $v$ allows order comparison for special factors in the words $u$ and $v$.

**Lemma 8** *Let $u,v \in A^+$ be words and suppose $u = x_1 u_1 \cdots x_k u_k = u'_1 y_1 \cdots u'_\ell y_\ell$ and $v = x_1 v_1 \cdots x_k v_k = v'_1 y_1 \cdots v'_\ell y_\ell$ with $k,\ell \geq 1$, $u'_1, v'_1, u_k, v_k \in A^*$ and $x_i, y_i \in A^+$ are factorizations such that*

- *$x_1 u_1 \cdots x_k$ is the shortest prefix of $u$ contained in $x_1 A^+ x_2 \cdots A^+ x_k$ and $x_1 v_1 \cdots x_k$ is the shortest prefix of $v$ contained in $x_1 A^+ x_2 \cdots A^+ x_k$,*
- *$y_1 \cdots u'_\ell y_\ell$ is the shortest suffix of $u$ contained in $y_1 A^+ y_2 \cdots A^+ y_\ell$ and $y_1 \cdots v'_\ell y_\ell$ is the shortest suffix of $v$ contained in $y_1 A^+ y_2 \cdots A^+ y_\ell$.*

*Let $\Delta_u = |u| - |x_1 u_1 \cdots u_{k-1}| - |u'_2 \cdots u'_\ell y_\ell|$ and let $\Delta_v = |v| - |x_1 v_1 \cdots v_{k-1}| - |v'_2 \cdots v'_\ell y_\ell|$. If $u \approx_{1,n} v$ for $n = |x_1 \cdots x_k| + |y_1 \cdots y_\ell|$, then the relative order of the occurrences of $x_k$ and $y_1$ is the same in $u$ and $v$, i.e., one of the following conditions applies:*

*1) $\Delta_u > |x_k y_1|$ and $\Delta_v > |x_k y_1|$.*
*2) $\Delta_u < 0$ and $\Delta_v < 0$.*
*3) $\Delta_u = \Delta_v$.*

*Proof:* Suppose that $\Delta_u > |x_k y_1|$, i.e., in $u$ all positions of the occurrence of $y_1$ are at least by one greater than all positions of the occurrence of $x_k$. Then $u$ is contained in the monomial $P = x_1 A^+ \cdots x_k A^+ y_1 A^+ \cdots y_\ell$. By assumption, $v$ is also contained in $P$ and thus $\Delta_v > |x_k y_1|$. Symmetry yields $\Delta_u > |x_k y_1|$ if and only if $\Delta_v > |x_k y_1|$.

Suppose that $0 \leq \Delta_u \leq |x_k y_1|$, i.e., in $u$ the occurrences of $x_k$ and $y_1$ are adjacent or overlap each other. By the above, we can assume $\Delta_v \leq |x_k y_1|$. Let $1 \leq i \leq k$ be minimal such that $|x_1 u_1 \cdots x_i| \geq |u'_1|$ and let $1 \leq j \leq \ell$ be maximal such that $|u'_1 y_1 \cdots u'_j| \leq |x_1 u_1 \cdots x_k|$. Let $p$ be the shorter of the two prefixes $u'_1$ and $x_1 u_1 \cdots x_{i-1} u_{i-1}$ of $u$. Similarly, let $q$ be the shorter of the two suffixes $u_k$ and $u'_{j+1} y_{j+1} \cdots u'_\ell y_\ell$. Then $z = p^{-1} u q^{-1}$ is the factor of $u$ corresponding to all positions of $x_m$'s and $y_m$'s which are adjacent to or overlapping with $x_k$ or $y_1$. We have that $u$ is contained in $x_1 A^+ \cdots x_{i-1} A^+ z A^+ y_{j+1} \cdots A^+ y_\ell$ and by assumption so is $v$. Note that $|x_1 \cdots x_{i-1} z y_{j+1} \cdots y_\ell| \leq n$. Hence, relative to the occurrence of $y_1$, the occurrence of $x_k$ cannot be further to the right in $v$ than in $u$, i.e., $0 \leq \Delta_u \leq \Delta_v \leq |x_k y_1|$. By symmetry $0 \leq \Delta_u \leq |x_k y_1|$ if and only if $0 \leq \Delta_v \leq |x_k y_1|$; and if this is the case, then $\Delta_u = \Delta_v$.

By the above we also get $\Delta_u < 0$ if and only if $\Delta_v < 0$. This shows that at least one of the conditions (1), (2), or (3) applies. □

The main combinatorial ingredient to go from $\mathbf{W}_m$ to $\mathrm{FO}_m^2$ is the factorization given in the following lemma. It combines our findings so far regarding **LDA** and $\approx_{m,n}$.

**Lemma 9** *Let $S \in$ **LDA**, let $m \geq 2$, let $N = 2|S|^2$, and let $u, v \in S^+$ with $u \approx_{m,n+N} v$. There exist factorizations $u = w_0 s_1 w_1 \cdots s_\ell w_\ell$ and $v = w_0 t_1 w_1 \cdots t_\ell w_\ell$ with $w_i, s_i, t_i \in S^+$, and $|w_0 \cdots w_\ell| \leq N$ such that for all $1 \leq i \leq \ell$ the following hold:*
*1) $s_i \approx_{m-1,n} t_i$,*
*2) $w_0 s_1 \cdots w_{i-1} \mathcal{R} w_0 s_1 \cdots w_{i-1} s_i$ in $S$,*
*3) $w_i \cdots t_\ell w_\ell \mathcal{L} t_i w_i \cdots t_\ell w_\ell$ in $S$.*

*Proof:* Let $X' = \{1\} \cup \{i \in \mathbb{N} \mid 1 < i \leq |u|, u[1; i-1] >_\mathcal{R} u[1;i] \text{ in } S\}$ be the set of positions of $u$ which cause an $\mathcal{R}$-descent when reading from left to right. Let $X$ be the set of positions $j$ such that there exists $i \in X'$ with $0 \leq i - j \leq |S|$, i.e., we include all $|S|$ positions to the left of each $i \in X'$. Let $Y'$ and $Y$ be defined left-right symmetrically on $v$, i.e., $Y' = \{|v|\} \cup \{i \in \mathbb{N} \mid 1 < i \leq |v|, v[i-1;|v|] >_\mathcal{L} u[i;|v|] \text{ in } S\}$ and $Y$ is the set of positions $j$ such that $0 \leq j - i \leq |S|$ for some $i \in Y'$. Let $X = X_1 \cup \cdots \cup X_k$ where $X_i \neq \emptyset$ are maximal subsets of consecutive positions of $X$ such that all positions of $X_i$ are smaller than all positions of $X_{i+1}$. Symmetrically, let $Y = Y_1 \cup \cdots \cup Y_{k'}$ where $Y_i \neq \emptyset$ are maximal subsets of consecutive positions of $Y$ and such that all positions of $Y_i$ are smaller than all positions of $Y_{i+1}$.

Let $x_i = u[X_i]$ and $y_i = u[Y_i]$ be the factors of $u$ and $v$ covered by the positions of $X_i$ and $Y_i$, respectively. By construction and Lemma 4 (1), we see that $u[1; \max(X_i)]$ is the shortest prefix of $u$ which is contained in $x_1 S^+ x_2 \cdots S^+ x_i$. Symmetrically, $v[\min(Y_i); |v|]$ is the shortest suffix of $v$ which is contained in $y_i S^+ y_{i+1} \cdots S^+ y_{k'}$ by Lemma 4 (2). We use these properties to transfer the positions of $X$ to $v$ and the positions of $Y$ to $u$. More precisely, we let $Y'' = Y_1'' \cup \cdots \cup Y_{k'}''$ be such that each $Y_i''$ is an interval of positions of $u$ with $u[Y_i''] = y_i$ and $u[\min(Y_i''); |u|]$ is the shortest suffix of $u$ which is contained in $y_i S^+ y_{i+1} \cdots S^+ y_{k'}$. And we let $X = X_1'' \cup \cdots \cup X_k''$ be such that each $X_i''$ is an interval of positions of $v$ with $v[X_i''] = x_i$ and $v[1; \max(X_i'')]$ is the shortest prefix of $v$ which is contained in $x_1 S^+ x_2 \cdots S^+ x_i$. Note that $u \in S^* y_1 S^+ y_2 \cdots S^+ y_{k'}$ and $v \in x_1 S^+ x_2 \cdots S^+ x_k S^*$ since $u \approx_{m,n+N} v$.

Now, consider the factorization $u = w_0 s_1 w_1 \cdots s_\ell w_\ell$ with $s_i \in S^+$ such that the $w_i$ are the factors covered by maximal subsets of consecutive positions of $X \cup Y''$. Intuitively, this means that we merge overlapping and adjacent factors $x_i$ and $y_j$ in $u$. Lemma 8 shows that the relative order of those concrete occurrences of $x_i$ and $y_j$ in $u$ and in $v$ is the same. Therefore, if we consider the factorization of $v$ which is covered by maximal subsets of consecutive positions in $X'' \cup Y$, then we end up with the same factors in the same order, i.e., we have $v = w_0 t_1 w_1 \cdots t_\ell w_\ell$ for some $t_i \in S^+$.

Since the $\mathcal{R}$-class and the $\mathcal{L}$-class can descend at most $|S| - 1$ times, we have $|X' \cup Y'| \leq 2|S|$ and thus $|w_0 \cdots w_\ell| \leq |X \cup Y''| \leq 2|S|^2$. Moreover, by construction, every $\mathcal{R}$-descent when reading prefixes of $u$ as well as every $\mathcal{L}$-descent when reading suffixes of $v$ is covered by some factor $w_i$ showing (2) and (3).

It remains to show $s_i \approx_{m-1,n} t_i$ for all $i$. An intermediate step is the following claim.

**Claim** *If $s_k w_k \cdots s_\ell w_\ell \approx_{m,n+N} t_k w_k \cdots t_\ell w_\ell$ for some $N \geq |w_k \cdots w_\ell|$, then $s_i \approx_{m-1,n} t_i$ for all $i \in \{k, \ldots, \ell\}$.*

The proof of this claim is by induction on $\ell - k$. Every $w_i$ either arises from some $x_j$'s or some $y_j$'s or both. Therefore, the $w_i$'s inherit the properties of the corresponding $x_j$'s and $y_j$'s of being the first occurrence (respectively being the last occurrence). If there is no $w_i$ arising from an $x_j$, then every $w_i$ has a unique occurrence in $w_i s_{i+1}$ as well as in $w_i t_{i+1}$. Thus $s_i \approx_{m-1,n} t_i$ for all $i$ by an $(\ell - k)$-fold application of condition (2) in the definition of $\approx_{m,n}$ (from right to left). For $i = k$ this uses Lemma 6.

Fix the first $w_i$ which arises from an $x_j$. We have $s_j \approx_{m-1,n} t_j$ for all $j > i$ by condition (1) in the definition of $\approx_{m,n}$ and induction. If $i = k$, then $s_k \approx_{m-1,n} t_k$ again by condition (1) in the definition of $\approx_{m,n}$. Assume therefore $i > k$ in the sequel. Let $h \geq i$ be minimal such that $w_h$ arises from some $y_j$; note that $w_\ell$ arises from $y_{k'}$. By a repeated application of condition (2) in the definition of $\approx_{m,n}$ we get that $s_k w_k \cdots s_h \approx_{m,n+N'} t_k w_k \cdots t_h$ for $N' = |w_k \cdots w_{h-1}|$. Now, $w_{i-1}$ has a unique occurrence in each of the words $w_{i-1} s_i \cdots s_h$ and $w_{i-1} t_i \cdots t_h$. Therefore, by repeatedly applying condition (2) in the definition of $\approx_{m,n}$ we see that $s_j \approx_{m-1,n} t_j$ for all $k \leq j < i$. If $h > i$, then by condition (3) in the definition of $\approx_{m,n}$ we see that $s_i \approx_{m-1,n} t_i$; and if $h = i$, then this follows from condition (2) in the definition of $\approx_{m,n}$. This concludes the proof of the claim.

Now by condition (1) in the definition of $\approx_{m,n}$, we see $s_1 w_1 \cdots s_\ell w_\ell \approx_{m,n+N'} t_1 w_1 \cdots t_\ell w_\ell$ for $N' = N - |w_0|$ and the above claim yields $s_j \approx_{m-1,n} t_j$ for all $1 \leq j \leq \ell$. □

The following proposition essentially shows how to pass from $\mathbf{W}_m$ to $\mathrm{FO}^2_m[<, \mathrm{suc}]$. The key to its proof is a string rewriting system which enables induction on the parameter $m$. Intuitively, for $S \in \mathbf{W}_m$ we consider the maximal quotient contained in $\mathbf{W}_{m-1}$. Since the latter is given by an omega-identity, this quotient can be described by a string rewriting system. A single rewriting step of this system corresponds to one application of the omega-identity for $\mathbf{W}_{m-1}$ and can be lifted to $\mathbf{W}_m$ relatively easily.

**Proposition 1** *For every $S \in \mathbf{W}_m$ with $m \geq 1$ there exists $n \geq 1$ such that $u \approx_{m,n} v$ implies $u = v$ in $S$ for all $u, v \in S^+$.*

*Proof:* We perform an induction on $m$. By Knast's Theorem [9], if $L$ is recognized by a semigroup in $S \in \mathbf{W}_1$, then the language $L$ is a Boolean combination of monomials $w_1 A^+ w_2 \cdots A^+ w_\ell$. Choosing $n \geq 1$ such that for all these monomials we have $|w_1 \cdots w_\ell| \leq n$ yields the claim for $m = 1$.

Let $\omega > |S|$ be an integer such $x^\omega$ is idempotent in $S$ for all $x \in S$. Consider the relation $\to$ on $S^+$ given by $s \to t$ if $s = t$ in $S$ or if $s = p u_{m-1} q$ and $t = p v_{m-1} q$ for some $p, q \in S^*$ and some $x_i, e, y_i, f, p_i, q_i, z, z' \in S^+$ such that $u_1 = (e^\omega z f^\omega x_1 e^\omega)^\omega z (f^\omega y_1 e^\omega z' f^\omega)^\omega$ and $v_1 = (e^\omega z f^\omega x_1 e^\omega)^\omega z' (f^\omega y_1 e^\omega z' f^\omega)^\omega$ and for $i \geq 2$ we have

$$u_i = (p_i u_{i-1} q_i x_i)^\omega p_i u_{i-1} q_i (y_i p_i u_{i-1} q_i)^\omega, \qquad v_i = (p_i u_{i-1} q_i x_i)^\omega p_i v_{i-1} q_i (y_i p_i u_{i-1} q_i)^\omega.$$

Let $\stackrel{*}{\leftrightarrow}$ be the reflexive, symmetric and transitive closure of $\to$. The relation $\stackrel{*}{\leftrightarrow}$ is a congruence of finite index (since $S^+/\stackrel{*}{\leftrightarrow}$ is a quotient of $S$). Moreover, $x^\omega \stackrel{*}{\leftrightarrow} x^{2\omega}$ for all $x \in S^+$ and $S^+/\stackrel{*}{\leftrightarrow} \in \mathbf{W}_{m-1}$.

**Claim 1** *Let $u, s, t \in S^+$. If $s \to t$, then $u \mathcal{R} us$ in $S$ if and only if $u \mathcal{R} ut$ in $S$.*

Assume without restriction that $s \neq t$ in $S$. By construction of $u_{m-1}$ and $v_{m-1}$, we have $\mathrm{alph}_{|S|+1}(s) = \mathrm{alph}_{|S|+1}(t)$. Note that by choice of $\omega$, in particular both words have the same prefix and the same suffix of length $|S|+1$. Lemma 3 yields the claim.

**Claim 2** *Let $u, v, s, t \in S^+$ with $s \overset{*}{\leftrightarrow} t$. If $u \mathcal{R} us$ and $v \mathcal{L} tv$ in $S$, then $usv = utv$ in $S$.*

Since $s \overset{*}{\leftrightarrow} t$, there exists $k \geq 0$ and $w_0, \ldots, w_k \in S^+$ such that $s = w_0$ and $w_k = t$ and such that either $w_{i-1} \to w_i$ or $w_i \to w_{i-1}$ for each $1 \leq i \leq k$. Claim 1 and its left-right dual, yield that $u \mathcal{R} uw_i$ and $v \mathcal{L} w_i v$ in $S$ for all $i$. It therefore suffices to show the claim for $s \to t$. The claim is trivial if $s = t$ in $S$. Otherwise suppose $s = p_m u_{m-1} q_m$ and $t = p_m v_{m-1} q_m$. Since $u \mathcal{R} us$ in $S$, there exists $x_m \in S$ such that $u = us x_m$ in $S$. Since $v \mathcal{L} tv$ in $S$ the left-right dual of Claim 1 implies $v \mathcal{L} sv$ in $S$. Hence, there exists $y_m \in S$ such that $v = y_m s v$ in $S$. Now, $u = u(p_m u_{m-1} q_m x_m)^\omega$ in $S$ and $v = (y_m p_m u_{m-1} q_m)^\omega v$ in $S$ and with $S \in \mathbf{W}_m$ we see

$$usv = u(p_m u_{m-1} q_m x_m)^\omega p_m u_{m-1} q_m (y_m p_m u_{m-1} q_m)^\omega v$$
$$= u(p_m u_{m-1} q_m x_m)^\omega p_m v_{m-1} q_m (y_m p_m u_{m-1} q_m)^\omega v = utv \text{ in } S,$$

thus establishing the claim.

Since $S^+/\overset{*}{\leftrightarrow} \in \mathbf{W}_{m-1}$, by induction there exists $n \geq 1$ such that $s \approx_{m-1,n} t$ implies $s \overset{*}{\leftrightarrow} t$ for all $s, t \in S^+$. Let $u, v \in S^+$ and suppose $u \approx_{m,n+N} v$ for $N = 2|S|^2$. Let $u = w_0 s_1 w_1 \cdots s_\ell w_\ell$ and $v = w_0 t_1 w_1 \cdots t_\ell w_\ell$ be the factorizations given by Lemma 9; in particular $s_i \approx_{m-1,n} t_i$ and $w_0 s_1 \cdots w_{i-1} \mathcal{R} w_0 s_1 \cdots w_{i-1} s_i$ in $S$ and $w_i \cdots t_\ell w_\ell \mathcal{L} t_i w_i \cdots t_\ell w_\ell$. By choice of $n$ we have $s_i \overset{*}{\leftrightarrow} t_i$ for all $i$ and repeated application of Claim 2 yields the following chain of identities valid in $S$:

$$v = w_0 t_1 w_1 t_2 \cdots t_{\ell-1} w_{\ell-1} t_\ell w_\ell$$
$$= w_0 s_1 w_1 t_2 \cdots t_{\ell-1} w_{\ell-1} t_\ell w_\ell$$
$$= w_0 s_1 w_1 s_2 \cdots t_{\ell-1} w_{\ell-1} t_\ell w_\ell$$
$$\vdots$$
$$= w_0 s_1 w_1 s_2 \cdots s_{\ell-1} w_{\ell-1} t_\ell w_\ell$$
$$= w_0 s_1 w_1 s_2 \cdots s_{\ell-1} w_{\ell-1} s_\ell w_\ell = u.$$

This concludes the proof. □

### 3.2 From First-Order Logic to Algebra

Next, we shall show the reverse direction, *i.e.*, languages definable in $\mathrm{FO}_m^2$ with successor are recognizable in $\mathbf{W}_m$. Our proof is inspired by Straubing's proof showing that $\mathrm{FO}_m^2[<]$-definable languages are recognized in the so-called weakly iterated two-sided semidirect product $((\mathbf{J} ** \mathbf{J}) ** \mathbf{J}) \cdots ** \mathbf{J}$ where $\mathbf{J}$ appears $n$ times [26]. However, to avoid technical notation we do not use semidirect products.

An intermediate step towards this is the following lemma reducing the alternation depth by encoding some information into an extended alphabet. To formalize this, for $n \geq 1$ consider the length-preserving map $\alpha_n$ whose $x^{\text{th}}$ letter of the image of $u$ is given by $\big(\mathcal{P}_n(u[1; x-1]), a, \mathcal{P}_n(u[x+1; |u|])\big)$ where $a$ is the $x^{\text{th}}$ letter of $u$ and where for $v \in A^*$

$$\mathcal{P}_n(v) = \big\{(w_1, \ldots, w_\ell) \mid v \in w_1 A^+ w_2 \cdots A^+ w_\ell,\ w_i \in A^*,\ |w_1 \cdots w_\ell| \leq n,\ 1 \leq \ell \leq n\big\}.$$

Note that $\mathcal{P}_n(v) = \{(1)\}$ if and only if $v = 1$. We view $\alpha_n(u)$ as a word over the alphabet $2^B \times A \times 2^B$ with $B = \{(w_1, \ldots, w_\ell) \mid w_i \in A^*, |w_1 \cdots w_\ell| \leq n, 1 \leq \ell \leq n\}$. Note however that $\alpha_n$ is *not* a homomorphism.

**Lemma 10** *For all $\varphi(x) \in \mathrm{FO}^2_{1,n}[<, \mathrm{suc}]$ with free variable $x$ there exists a formula $\varphi'(x) \in \mathrm{FO}^2_{0,0}[<, \mathrm{suc}]$ such that for all $u \in A^+$ and all positions $x \in \mathrm{pos}(u)$ we have*

$$u \models \varphi(x) \quad \text{if and only if} \quad \alpha_n(u) \models \varphi'(x).$$

*For all $m \geq 2$ and all $\varphi \in \mathrm{FO}^2_{m,n}[<, \mathrm{suc}]$ there exists a formula $\varphi' \in \mathrm{FO}^2_{m-1,n}[<, \mathrm{suc}]$ such that for all $u \in A^+$ we have*

$$u \models \varphi \quad \text{if and only if} \quad \alpha_n(u) \models \varphi'.$$

*Proof:* First, suppose that $\varphi(x)$ is a purely existential formula in $\mathrm{FO}^2[<, \mathrm{suc}]$ with quantifier depth $n$. Using the standard procedure (which introduces new variables), we get an equivalent formula with quantifier depth $n$ in prenex normal form, *i.e.*,

$$\varphi(x) \equiv \exists x_1 \cdots \exists x_n \, \varphi'$$

where $\varphi' = \varphi'(x, x_1, \ldots, x_n)$ is quantifier-free.

In the next step, we make explicit the order type and the label of the variables. To do this concisely, we introduce the following vector notation. We denote by $\underline{z} = (z_1, \ldots, z_k)$ a vector of variables and set $|\underline{z}| = k$. Let $\exists \underline{z}$ be a shortcut for $\exists z_1 \cdots \exists z_k$, let $\lambda(\underline{z}) = w$ be a shortcut which states that $z_1 < \cdots < z_k$ are consecutive positions labeled by the word $w$. The shortcut $\underline{z} < \underline{y}$ requires that all positions of $\underline{z}$ be smaller than all positions of $\underline{y}$, and the shortcut $\neg\mathrm{suc}(\underline{z}, \underline{y})$ states that no position of $\underline{y}$ is the successor of any position of $\underline{z}$. For a fixed ordering and labeling of the variables, the truth of all atomic formulae and thus also of $\varphi'$ is completely determined. We now consider all possibilities to order and label the variables such that $\varphi'$ is true. Reordering variables and congregating consecutive variables to vectors we end up with a finite disjunction of formulae $\psi(x)$ of the form

$$\exists \underline{z}_1 < x \cdots \exists \underline{z}_k < x \, \exists \underline{z}_{k+1} > x \cdots \exists \underline{z}_\ell > x \colon (\lambda(x) = a) \wedge \bigwedge_i (\underline{z}_{i-1} \ll \underline{z}_i \wedge (\lambda(z_i) = w_i)) \wedge \chi$$

for some $w_i \in A^+$, some $a \in A$ and some $\chi$ of the form $\chi_1 \wedge \chi_2$ where

$$\chi_1 \in \{\mathrm{suc}(\underline{z}_k, x), \neg\mathrm{suc}(\underline{z}_k, x)\}, \quad \chi_2 \in \{\mathrm{suc}(x, \underline{z}_{k+1}), \neg\mathrm{suc}(x, \underline{z}_{k+1})\}.$$

The variable $x$ separates $z_i$ from $z_j$ for $i \leq k < j$. This allows to split $\psi(x)$ at index $i$ to get conjunction of $\lambda(x) = a$, of formulae of the form

$$\varsigma(x) \equiv \exists \underline{z}_1 < x \cdots \exists \underline{z}_k < x \colon \bigwedge_i (\underline{z}_{i-1} \ll \underline{z}_i \wedge (\lambda(z_i) = w_i)) \wedge \chi_1$$

and of similar formulae $\varsigma'(x)$ involving the variables $\underline{z}_j$ with $j > k$.

Now, for every model $u \in A^+$ every position $x \in \mathrm{pos}(u)$ the truth of $u \models \varsigma(x)$ is completely determined by the set $\mathcal{P}_n(u[1; x-1])$. If for example $\chi_1 \equiv \mathrm{suc}(\underline{z}_k, x)$, then $u \models \varsigma(x)$ if and only if $u[1; x-1] \in A^* w_1 \cdots A^+ w_k$ if and only if $\{(1, w_1, \ldots, w_k), (w_1, \ldots, w_k)\} \cap \mathcal{P}_n(u[1; x-1]) \neq \emptyset$. A symmetric argument shows that $\varsigma'(x)$ is completely determined by the set $\mathcal{P}_n(u[x+1; |u|])$. Therefore, $u \models \psi(x)$ is completely determined by the $x^\mathrm{th}$ letter of $\alpha_n(u)$, *i.e.*, there exists a formula $\psi'(x)$ which is a Boolean combination of labels over the alphabet $2^B \times A \times 2^B$ such that $u \models \psi(x)$ if and only if $\alpha_n(u) \models \psi'(x)$. This implies the first part for purely existential formulae. For a purely universal formula $\varphi(x)$, we consider its negation and use the equivalence $\neg(\forall x\, \xi) \equiv \exists x\, \neg \xi$ and De Morgan's laws to move the negation inwards so that no negation ranges over any quantifier. Let $\psi(x)$ be the resulting purely existential formula. Setting $\varphi'(x) \equiv \neg \psi'(x)$ yields the claim. The extension to Boolean combinations of quantifier free formulae is straightforward. This concludes the first part of the proof.

For the second part of the claim let $m \geq 2$ and $\varphi \in \mathrm{FO}^2_{m,n}[<, \mathrm{suc}]$. Without restriction, we may assume that no negation ranges over any quantifier. Consider an *innermost quantified block* of $\varphi$, i.e., a maximal subformula $\psi$ of $\varphi$ of the form $\exists y\, \xi$ or of the form $\forall y\, \xi$ where all quantifiers in $\psi$ are of the same type. Note that $y$ does not appear freely in $\psi$ and thus $\psi = \psi(x)$ is a formula with one free variable $x$. We replace every innermost quantified block $\psi$ by the formula $\psi'$ of the first part of the lemma. Note that we indeed have $\psi \in \mathrm{FO}^2_{1,n}[<, \mathrm{suc}]$. This decreases the alternation depth by one since the $\psi$ are maximal. It might happen that the resulting formulae contains label predicates over the alphabet $A$. The formulae $\varphi'$ is obtained by replacing $\lambda(x) = a$ with $a \in A$ by the disjunction of formulae $\lambda(x) = (P, a, Q)$ ranging over all $P, Q \subseteq B$. $\square$

The next lemma shows that for any given language definable in $\mathrm{FO}^2_m[<, \mathrm{suc}]$ there is a recognizing semigroup in $\mathbf{W}_m$. Let $u \cong_{m,n} v$ over $A^+$ if $puq \models \varphi \Leftrightarrow pvq \models \varphi$ for all $p, q \in A^*$ and all $\varphi \in \mathrm{FO}^2_{m,n}[<, \mathrm{suc}]$.

**Lemma 11** *Let $m, n \geq 1$. For $2 \leq i \leq m$ define the following words over the alphabet $\Sigma = \{x_i, y_i, p_i, q_i, e, f, s, t \mid 1 \leq i \leq m\}$:*

$$U_1 = p_1(esfx_1e)^n\, s\, (fy_1etf)^n q_1, \qquad U_i = p_i(U_{i-1}x_i)^n U_{i-1}(y_i U_{i-1})^n q_i,$$
$$V_1 = p_1(esfx_1e)^n\, t\, (fy_1etf)^n q_1, \qquad V_i = p_i(U_{i-1}x_i)^n V_{i-1}(y_i U_{i-1})^n q_i.$$

*Then for all homomorphisms $h : \Sigma^+ \to A^+$ such that $h(e) \cong_{m,n} h(e^2)$ and $h(f) \cong_{m,n} h(f^2)$ we have $h(U_m) \cong_{m,n} h(V_m)$. In particular we have $A^+/\cong_{m,n} \in \mathbf{W}_m$.*

*Proof:* In principle, the claim is that $\mathrm{FO}^2_m[<, \mathrm{suc}]$ with order and successor is unable to disprove the defining identity of $\mathbf{W}_m$. For conciseness, we use a slightly modified version of this identity. The homomorphism $h$ gives a valuation of the variables in $\Sigma$.

Note that it suffices to show $h(U_m) \models \varphi \Leftrightarrow h(V_m) \models \varphi$ all $\varphi \in \mathrm{FO}^2_{m,n}[<, \mathrm{suc}]$ and all $h$. By incorporating $p$ into the image of $p_m$ and $q$ into the image of $q_m$, this indeed implies $ph(U_m)q \models \varphi \Leftrightarrow ph(V_m)q \models \varphi$ for all $p, q \in A^*$.

We perform an induction with $m = 1$ as base. Consider a sentence $\varphi \in \mathrm{FO}^2_{1,n}[<, \mathrm{suc}]$. Even though $\varphi$ is a sentence, we may view it as a formula $\varphi(x)$ in one free variable $x$ whose truth does not depend on the interpretation of $x$. Let $\varphi'(x) \in \mathrm{FO}^2_{0,0}$ be the formula from Lemma 10. Then we have $h(U_1) \models \varphi$ if and only if $\alpha_n(h(U_1)) \models \varphi'(1)$ if and only if $\alpha_n(h(V_1)) \models \varphi'(1)$ if and only if $h(V_1) \models \varphi$. The second equivalence holds since $h(U_1)$ and $h(V_1)$ have the same short sequences of short factors, i.e., $\mathcal{P}_n(h(U_1)) = \mathcal{P}_n(h(V_1))$.

Let $m \geq 2$ and consider $U_m, V_m$ and a valuation $h : \Sigma^+ \to A^+$ as defined in the statement of the lemma. Expanding the definition yields the following structure of $U_m$ and $V_m$:

$$U_m = p_m(U_{m-1}x_m)^n \cdots p_2(U_1 x_2)^n\, p_1(esfx_1e)^n s(fy_1etf)^n q_1\, (y_2 U_1)^n q_2 \cdots (y_m U_{m-1})^n q_m,$$
$$V_m = p_m(U_{m-1}x_m)^n \cdots p_2(U_1 x_2)^n\, p_1(esfx_1e)^n t(fy_1etf)^n q_1\, (y_2 U_1)^n q_2 \cdots (y_m U_{m-1})^n q_m.$$

Since $h(e)$ and $h(f)$ are idempotent in $A^+/\cong_{m,n}$, we may assume without restriction that

- $h(e)$ and $h(f)$ have lengths at least $n$, and

- $h(e)$ is a prefix of $h(s)$ and of $h(t)$ and a suffix of $h(p_1)$, of $h(x_1)$ and of $h(y_1)$, and
- $h(f)$ is a suffix of $h(s)$ and of $h(t)$ and a prefix of $h(q_1)$, of $h(x_1)$ and of $h(y_1)$.

Let $U'_m = \alpha_n(h(U_m))$ and $V'_m = \alpha_n(h(V_m))$. We now want to show that the factors of $U'_m$ and $V'_m$ which come from the central factors $U_{m-1}$ (in $U_m$) and $V_{m-1}$ (in $V_m$) can be obtained as the image of $U_{m-1}$ and $V_{m-1}$ under some valuation $h' : \Sigma^+ \to C^+$ where $C = 2^B \times A \times 2^B$.

Let $h(U_m) = pq$ with $p$ being a prefix of $h\bigl(p_m(U_{m-1}x_m)^n \cdots p_2(U_1 x_2)^n p_1(esfx_1e)^n\bigr)$. Then $h(V_m) = pq'$ for some $q'$ and we have $\mathcal{P}_n(q) = \mathcal{P}_n(q')$. Symmetrically, if $h(U_m) = pq$ for some suffix $q$ of $h\bigl((fy_1etf)^n q_1(y_2 U_1)^n q_2 \cdots (y_m U_{m-1})^n q_m\bigr)$, then $h(V_m) = p'q$ for some $p'$ and $\mathcal{P}_n(p) = \mathcal{P}_n(p')$. This in particular implies that $U'_m$ and $V'_m$ have the same prefix $p$ of length $|h(p_m(U_{m-1}x_m)^n)|$ and the same suffix $q$ of length $|h((y_m U_{m-1})^n q_m)|$. Let $U'_{m-1} = p^{-1} U'_m q^{-1}$ and $V'_{m-1} = p^{-1} V'_m q^{-1}$. We define a homomorphism $h' : \Sigma^+ \to C^+$ by setting for $z \in \Sigma$

$$h'(z) = C^{-|h(p')|} \, U'_m \, C^{-|h(q')|}$$

if $p', q' \in \Sigma^*$ are such that $U_m = p'zq'$ and $|p'| \geq |p_m(U_{m-1}x_m)^n|$ and $|q'| \geq |(y_m U_{m-1})^n q_m|$. In other words, we take an *arbitrary* occurrence of $z$ in the central part of $U_m$ and define the image $h'(z)$ as the factor of $U'_m$ which corresponds to this occurrence. We now show that this mapping is well-defined, *i.e.*, the value $h'(z)$ does not depend on the concrete location of $z$ chosen.

Suppose $W = U_m$ or $W = V_m$ and $W_m = p''zq''$ for some $p'', q''$ with $|h(p'')| \geq |p_m(U_{m-1}x_m)^n|$ and $|h(q'')| \geq |(y_m U_{m-1})^n q_m|$ is another occurrence of $z$ in the central part of $U_m$ or $V_m$. By the assumptions on the valuation $h$, both occurrences of $z$ have the same surrounding of $n$ letters under the valuation $h$, *i.e.*, $h(p'), h(p'')$ have the same suffix of length $n$, and $h(q'), h(q'')$ have the same prefix of length $n$. Let $P = A^* w_1 A^+ w_2 \cdots A^+ w_n A^*$ with $w_i \in A^+$ and $|w_1 \cdots w_\ell| \leq n$. Then $h(p') \in P$ if and only if $h(p_m(U_{m-1}x_m)^n) \in P$ if and only if $h(p'') \in P$; this holds because $h((U_{m-1}x_m)^n)$ contains short factors sufficiently often. Symmetrically $h(q') \in P$ if and only if $h(q'') \in P$. Moreover, $w_1 A^+ w_2 \cdots A^+ w_n = P \cap w_1 A^* \cap A^* w_n$. Thus if $h(z) = z_1 z_2$, then the sets $\mathcal{P}_n(h(p')z_1)$ and $\mathcal{P}_n(z_2 h(q'))$ do not depend on the concrete location of the variable $z$ given by $p'$ or $q'$, *i.e.*, we have

$$C^{-|h(p'')|} \, W \, C^{-|h(q'')|} = h'(z).$$

In other words, $\alpha_n$ annotates all occurrences of all variables within the central factors $U_{m-1}$ and $V_{m-1}$ consistently so that $U'_{m-1} = h'(U_{m-1})$ and $V'_{m-1} = h'(V_{m-1})$. Moreover, we may assume that $h'(e)$ and $h'(f)$ are idempotent in $C^+/\cong_{m-1,n}$. (Else let $\omega \geq 1$ be an integer such that $x^\omega$ is idempotent for all $x \in C^+/\cong_{m-1,n}$. And, instead of $h$, we consider the valuation $\hbar$ which maps $e$ to $h(e^\omega)$ and $f$ to $h(f^\omega)$ and all other $z \in \Sigma$ to $h(z)$. The construction of $\hbar'$ then yields $\hbar'(e) = h'(e)^\omega$.)

Induction yields $U'_{m-1} \cong_{m-1,n} V'_{m-1}$, and, because $\cong_{m-1,n}$ is a congruence, we see $U'_m \cong_{m-1,n} V'_m$. Consider a formula $\varphi \in \mathrm{FO}^2_{m,n}$ and let $\varphi' \in \mathrm{FO}^2_{m-1,n}$ be the formula from Lemma 10. Then $h(U_m) \models \varphi$ if and only if $U'_m \models \varphi'$ if and only if $V'_m \models \varphi'$ if and only if $h(V_m) \models \varphi$. This shows $h(U_m) \cong_{m,n} h(V_m)$. □

## 3.3 Temporal Logic and Completing the Proof

It remains to incorporate temporal logic into the picture. The following lemma is the hard part from $\mathrm{FO}^2[<, \mathrm{suc}]$ to $\mathrm{TL}_m[\mathsf{X}, \mathsf{F}, \mathsf{Y}, \mathsf{P}]$. The proof is done by a straight-forward syntactic translation. This can be also seen using a construction due to Etessami, Vardi, and Wilke [7, proof of Theorem 1] though their statement does not mention the alternation depth of the resulting temporal logic formula.

**Proposition 2** *Let $m \geq 1$ and $n \geq 0$. Every language definable in $\mathrm{FO}^2_{m,n}[<,\mathrm{suc}]$ is definable in $\mathrm{TL}_{m,3n}[\mathsf{X},\mathsf{F},\mathsf{Y},\mathsf{P}]$.*

*Proof:* The construction is by induction on the structure of the formula. For the inductive step, we also have to take free variables into account. Let $\varphi(x,y) \in \mathrm{FO}^2_{m,n}[<,\mathrm{suc}]$. We start by some normalizations on the structure of $\varphi$. We assume without restriction that on all paths in the parse tree of $\varphi$ no two successive quantifiers bind the same variable. Therefore, by starting the construction with a sentence having this property, we can assume that the first quantifier on every path in the parse tree of $\varphi$ binds the variable $y$, *i.e.*, when thinking of $\varphi$ as a subformula of some sentence, then on the path to $\varphi$ the previously bound variable is $x$. Moreover, we replace all universal quantifiers $\forall z\, \psi$ by $\neg \exists z\, \neg \psi$.

For integers $i, j$ let $\mathrm{ord}(i,j) \in \{\ll, -1, 0, +1, \gg\}$ be their *order-type* defined by

$$\mathrm{ord}(i,j) = \begin{cases} \ll & \text{if } i < j-1 \\ -1 & \text{if } i = j-1 \\ 0 & \text{if } i = j \\ +1 & \text{if } i = j+1 \\ \gg & \text{if } i > j+1. \end{cases}$$

Let $I \in \{\ll, -1, 0, +1, \gg\}$ and $a \in A$. We show that there exists a formula $\langle \varphi \rangle_{I,a}(x) \in \mathrm{TL}_{m,3n}[\mathsf{X},\mathsf{F},\mathsf{Y},\mathsf{P}]$ such that for all $u \in A^*$ and all positions $i, j \in \mathrm{pos}(u)$ with $u[j] = a$ and $\mathrm{ord}(i,j) = I$ we have $u \models \varphi(i,j)$ if and only if $u \models \langle \varphi \rangle_{I,a}(i)$.

The construction is by induction on the structure of the formula. Let $\langle \top \rangle_{I,a}(x) \equiv \top$, $\langle \bot \rangle_{I,a}(x) \equiv \bot$ and for the other atomic formulae we set

$$\langle \lambda(x) = b \rangle_{I,a}(x) \equiv b,$$

$$\langle \lambda(y) = b \rangle_{I,a}(x) \equiv \begin{cases} \top & \text{if } a = b, \\ \bot & \text{otherwise,} \end{cases}$$

$$\langle x < y \rangle_{I,a}(x) \equiv \begin{cases} \top & \text{if } I \in \{\ll, -1\}, \\ \bot & \text{otherwise,} \end{cases}$$

$$\langle \mathrm{suc}(x,y) \rangle_{I,a}(x) \equiv \begin{cases} \top & \text{if } I = -1, \\ \bot & \text{otherwise.} \end{cases}$$

The formulae $\langle y < x \rangle_{I,a}(x)$ and $\langle \mathrm{suc}(y,x) \rangle_{I,a}(x)$ are defined similarly. Conjunction is given by $\langle \varphi \wedge \psi \rangle_{I,a}(x) \equiv \langle \varphi \rangle_{I,a}(x) \wedge \langle \psi \rangle_{I,a}(x)$ and for disjunction we define $\langle \varphi \vee \psi \rangle_{I,a}(x) \equiv \langle \varphi \rangle_{I,a}(x) \vee \langle \psi \rangle_{I,a}(x)$. For negation we set $\langle \neg \varphi \rangle_{I,a}(x) \equiv \neg \langle \varphi \rangle_{I,a}(x)$. For existential quantification we let $\langle \exists y\colon \varphi \rangle_{I,a}(x)$ be given by

$$\bigvee_{b \in A} b \wedge \left( \mathsf{Y}\mathsf{Y}\mathsf{P} \langle \varphi \rangle_{\ll,b}(y) \vee \mathsf{Y} \langle \varphi \rangle_{-1,b}(y) \vee \langle \varphi \rangle_{0,b}(y) \vee \mathsf{X} \langle \varphi \rangle_{+1,b}(y) \vee \mathsf{X}\mathsf{X}\mathsf{F} \langle \varphi \rangle_{\gg,b}(y) \right).$$

In the construction of the formulae $\langle \varphi \rangle_{J,b}(y)$, the roles of $x$ and $y$ are interchanged. Note that due to our normalization at the beginning, we do not have to handle quantification over $x$.

Now, we inductively define $\langle\varphi\rangle \in \mathrm{TL}_{m,n}[\mathsf{X},\mathsf{F},\mathsf{Y},\mathsf{P}]$ for sentences $\varphi \in \mathrm{FO}^2_{m,n}[<,\mathrm{suc}]$. Boolean connectives are straightforward: $\langle\varphi \wedge \psi\rangle \equiv \langle\varphi\rangle \wedge \langle\psi\rangle$, $\langle\varphi \vee \psi\rangle \equiv \langle\varphi\rangle \vee \langle\psi\rangle$, and $\langle\neg\varphi\rangle \equiv \neg\langle\varphi\rangle$. For quantification we set $\langle\exists y\colon \varphi\rangle \equiv \mathsf{F}\,\langle\varphi\rangle_{I,a}(y)$ where $I \in \{\ll, -1, 0, +1, \gg\}$ and $a \in A$ are arbitrary. Hence if $L = L(\varphi)$ for some sentence $\varphi \in \mathrm{FO}^2_{m,n}[<,\mathrm{suc}]$, then $L = L(\langle\varphi\rangle)$ for $\langle\varphi\rangle \in \mathrm{TL}_{m,3n}[\mathsf{X},\mathsf{F},\mathsf{Y},\mathsf{P}]$. □

We are now ready to prove Theorem 1.

*Proof (Theorem 1):* We shall show "(1) ⇔ (2)" and "(1) ⇔ (3)".

"(1) ⇒ (2)": This follows by Proposition 2.

"(2) ⇒ (1)": This is trivial since the semantics of temporal logic formulae is given by two-variable first-order expressions.

"(1) ⇒ (3)": By Lemma 11 the semigroup $S = (A^+/\cong_{m,n})$ is in $\mathbf{W}_m$. Every language $L \subseteq A^+$ definable in $\mathrm{FO}^2_{m,n}[<,\mathrm{suc}]$ is a union of $\cong_{m,n}$-classes and hence recognized by $S$. The syntactic semigroup of $L$ is a quotient of $S$ and thus is also in $\mathbf{W}_m$.

"(3) ⇒ (1)": Suppose the homomorphism $h: A^+ \to S$ with $S \in \mathbf{W}_m$ recognizes $L \subseteq A^+$. Combining Proposition 1 and Lemma 7, we see that there exists an integer $n \geq 1$ such that $u \equiv_{m,n} v$ for $u,v \in S^+$ implies $u = v$ in $S$. Now, if $u \equiv_{m,n} v$ for $u,v \in A^+$, then $h(u) = h(v)$. Thus, by specifying the $\equiv_{m,n}$-classes of $A^+$ which are contained in $L$, we obtain a formula $\varphi \in \mathrm{FO}^2_{m,n}[<,\mathrm{suc}]$ such that $L(\varphi) = h^{-1}(h(L)) = L$. Note that the syntactic semigroup of $L$ recognizes $L$. □

## Conclusion

We showed that quantifier alternation for the logic $\mathrm{FO}^2[<,\mathrm{suc}]$ is decidable by giving a single identity of omega-terms for each level $\mathrm{FO}^2_m[<,\mathrm{suc}]$. The key ingredient in our proof is a rewriting technique which allows us to apply induction on $m$.

There is an algebraic construction $\mathbf{V} \mapsto \mathbf{V} * \mathbf{D}$ in terms of wreath products, see *e.g.* [25]. For most logical fragments $\mathcal{F}$, whenever $\mathcal{F}$ corresponds to a variety of finite monoids $\mathbf{V}$, then the fragment $\mathcal{F}'$ obtained from $\mathcal{F}$ by adding successor predicates corresponds to the semigroup variety $\mathbf{V} * \mathbf{D}$. This is also the case for $\mathrm{FO}^2_m[<]$ and $\mathrm{FO}^2_m[<,\mathrm{suc}]$. Therefore, if $\mathbf{V}_m$ is the variety of finite monoids corresponding to $\mathrm{FO}^2_m[<]$, then our result implies $\mathbf{V}_m * \mathbf{D} = \mathbf{W}_m$.

In general, decidability of $\mathbf{V}$ is not preserved by the operation $\mathbf{V} \mapsto \mathbf{V} * \mathbf{D}$, but a particularly nice situation occurs if $\mathbf{V} * \mathbf{D} = \mathbf{LV}$. Here, a semigroup $S$ is in $\mathbf{LV}$ if all local monoids of $S$ are in $\mathbf{V}$. For example the variety $\mathbf{DA}$ satisfies $\mathbf{DA} * \mathbf{D} = \mathbf{LDA}$, see [1, 4]. For $\mathbf{W}_1$ however, Knast has given an example showing $\mathbf{V}_1 * \mathbf{D} \neq \mathbf{LV}_1$. In view of this example, we conjecture that $\mathbf{V}_m * \mathbf{D} \neq \mathbf{LV}_m$ for all $m \geq 1$.

**Acknowledgments.** We would like to thank the anonymous referees of the conference version of this paper for many useful suggestions which helped to improve the presentation of the paper.